\documentclass[bibyear]{aa}

\usepackage{natbib}
\usepackage{graphicx}
\usepackage{lscape}
\usepackage{times}

\begin{document}

\title{Super-solar metallicity at the position of the ultra-long GRB~130925A\thanks{Based on observations taken at ESO/VLT, Programme ID 091.A-0703 and 092.A-0231}}

\author{P.~Schady\inst{\ref{inst1}}\and T.~Kr{\"u}hler\inst{\ref{inst2}}\and J.~Greiner\inst{\ref{inst1}}\and J.~F.~Graham\inst{\ref{inst1}}\and D.~A.~Kann\inst{\ref{inst3}}\and J.~Bolmer\inst{\ref{inst4},\ref{inst1}}\and C.~Delvaux\inst{\ref{inst1}}\and J.~Elliott\inst{\ref{inst5},\ref{inst1}}\and S.~Klose\inst{\ref{inst3}}\and F.~Knust\inst{\ref{inst1}}\and A.~Nicuesa~Guelbenzu\inst{\ref{inst3}}\and A.~Rau\inst{\ref{inst1}}\and A.~Rossi\inst{\ref{inst6}}\and S.~Savaglio\inst{\ref{inst7},\ref{inst1},\ref{inst8}}\and S.~Schmidl\inst{\ref{inst1}}\and T.~Schweyer\inst{\ref{inst4},\ref{inst1}}\and V.~Sudilovsky\inst{\ref{inst5}}\and M~.Tanga\inst{\ref{inst1}}\and N.~R.~Tanvir\inst{\ref{inst9}}\and K.~Varela\inst{\ref{inst1}}\and P.~Wiseman\inst{\ref{inst1}}}
\institute{Max-Planck-Institut f{\"u}r Extraterrestrische Physik, Giessenbachstra\ss e, 85748, Garching, Germany\\
\email{pschady@mpe.mpg.de}\label{inst1}
\and
European Southern Observatory, Alonso de C{\'o}rdova 3107, Vitacura, Casilla 19001 Santiago 19, Chile\label{inst2}
\and
Th{\"u}ringer Landessternwarte Tautenburg, Sternwarte 5, 07778 Tautenburg, Germany\label{inst3}
\and
Technische Universit{\"a}t M{\"u}nchen, Physik Dept., James-Franck-Stra\ss e, 85748 Garching, Germany\label{inst4}
\and
Harvard-Smithsonian Center for Astrophysics, 60 Garden Street, Cambridge, MA 02138, USA\label{inst5}
\and
INAF-IASF Bologna, Area della Ricerca CNR, via Gobetti 101, I--40129 Bologna, Italy\label{inst6}
\and
Physics Dept., University of Calabria, via P. Bucci, I-87036 Arcavacata di Rende, Italy\label{inst7}
\and
European Southern Observatory, Karl-Schwarzschild-Stra\ss e 2, 85748 Garching, Germany\label{inst8}
\and
Department of Physics and Astronomy, University of Leicester, University Road, Leicester LE1 7RH, UK\label{inst9}
}

\date{Received 9 March 2015 / Accepted 26 May 2015}

\abstract
{Over the last decade there has been immense progress in the follow-up of short and long gamma-ray bursts (GRBs), resulting in a significant rise in the detection rate of X-ray and optical afterglows, in the determination of GRB redshifts, and of the identification of the underlying host galaxies. Nevertheless, our theoretical understanding of the progenitors and central engines powering these vast explosions is lagging behind, and a newly identified class of ultra-long GRBs has fuelled speculation on the existence of a new channel of GRB formation. In this paper we present high signal-to-noise X-shooter observations of the host galaxy of GRB~130925A, which is the fourth unambiguously identified ultra-long GRB, with prompt $\gamma$-ray emission detected for $\sim 20$~ks. The GRB line of sight was close to the host galaxy nucleus, and our spectroscopic observations cover this region along the bulge/disk of the galaxy, and a bright star-forming region within the outskirts of the galaxy. From our broad wavelength coverage, we obtain accurate metallicity and dust-extinction measurements at the galaxy nucleus and at an outer star-forming region, and measure a super-solar metallicity at both locations, placing this galaxy within the $10-20$\% most metal-rich GRB host galaxies. Such a high metal enrichment has significant implications on the progenitor models of both long and ultra-long GRBs, although the edge-on orientation of the host galaxy does not allow us to rule out a large metallicity variation along our line of sight. The spatially resolved spectroscopic observations presented in this paper offer important insight into variations in the metal and dust abundance within GRB host galaxies. However, they also illustrate the need for integral field unit observations on a larger sample of GRB host galaxies of a variety of metallicities to provide a more quantitative view on the relation between the GRB circumburst environment and the galaxy-whole properties.}

\maketitle
\titlerunning{X-shooter observations of the host galaxy of GRB~130925A}
\authorrunning{Schady et al.}

\section{Introduction}
The collapsar model \citep{woo93,wm99,hmm05,yl05,wh06} has been largely successful at describing the overall observed properties of long-duration gamma-ray bursts (GRBs), and has been validated by the detection of supernovae coincident with the GRB explosions\footnotemark[1] (GRB~980425/SN1998bw; Galama et al., 1998, GRB030329/SN2003dh; Hjorth et al., 2003, Stanek et al., 2003). Nevertheless, the model in its simplest form has faced challenges; notably the recent rise in solar or super-solar metallicities measured in the environments of GRBs, which is in contention with the metallicity upper limit of $Z\mathrel{\hbox{\rlap{\hbox{\lower4pt\hbox{$\sim$}}}\hbox{$<$}}} 0.3Z_\odot$ imposed by the standard collapsar models \citep{woo93,mw99,wm99,hmm05,yl05,yln06,wh06}, and more recently, the detection of a possible new class of ultra-long duration GRB, which requires a longer-lasting central engine than can be provided by the Wolf-Rayet (WR) progenitors believed to produce standard long GRBs.
\footnotetext[1]{For the remainder of the paper we will use the term GRB to refer to the long-duration class of events unless specified otherwise.}

There have thus far been four clear examples of ultra-long duration GRBs, all with continuous prompt $\gamma$-ray emission lasting from thousands to tens of thousands of seconds rather than the more typical $10-100$s observed in standard long-duration GRBs. These are GRB~101225A \citep{tuf+11}, GRB~111209A \citep{sga+13,gsa+13,gmk+15}, GRB~121027A \citep{hgl+14}, and more recently, GRB~130925A, which had the longest duration prompt emission thus far detected in a GRB \citep[$\sim 20$~ks;][]{bbb+14,ewo+14,ptg+14,zs14}. The difficulty in accounting for the source of the long-lasting $\gamma$-ray emission, as well as notable differences in the X-ray light curve properties, have led to the suggestion that these ultra-long GRBs are a new class of transient, with progenitors that are distinct from those that produce standard long GRBs \citep{gsa+13,bgs15,lts+14,gmk+15}. However, other GRBs with evidence of long-lasting central engine activity, such as short periods of emission spread over long timescales ($\sim 1000$s), or long-lasting flaring at X-ray energies, may imply that there is a continuous distribution in the prompt emission duration of long GRBs, with the ultra-long class representing the extreme end of the distribution \citep[e.g.][]{vmp+13,zzm+14}. 

When considering those GRBs with continuous prompt emission for $>10^4$s, which we will refer to as the ultra-long class of events, the main progenitors explored have been tidal disruption events (TDE), magnetars, and low-metallicity blue supergiants (BSG) \citep{gsa+13,lts+14}. However, thus far the sample size of these events is too small to be able to strongly favour a single channel of formation. The unusually high host galaxy dust extinction measured along the line of sight to the most recently detected ultra-long GRB, GRB~130925A, introduces new diversity in the environmental properties associated with long, and especially ultra-long GRBs. With the typically rapid dissemination of sub-arcsecond GRB positions provided by the GRB-dedicated {\em Swift} mission \citep{gcg+04}, launched in 2004, and the commissioning of (semi-) robotic telescopes equipped with near-infrared (NIR) facilities, such as the GRB optical and near-infrared detector \citep[GROND;][]{gbc+08}, the Peters automated IR imaging telescope \citep[PAIRITEL;][]{bsb+06} and the reionization and transients IR camera \citep[RATIR;][]{bkf+12}, the detection of heavily dust-extinguished GRBs and dust-rich host galaxies has become more common. GRB~130925A was at a redshift of $z=0.347$ \citep{sks+13} and had a host galaxy visual extinction of A$_V$$=5.0\pm 0.7$~mag as measured from the afterglow spectral energy distribution (SED) \citep{gyk+14}, which is one of the largest host galaxy visual extinctions measured along the line of sight to any GRB, and certainly the largest along the line of sight to an ultra-long duration GRB.

The host galaxies of heavily dust extinguished GRBs (i.e. A$_V$$> 1$~mag) typically have larger stellar masses, luminosities and dust masses than the hosts of optically bright GRBs \citep{pcb+09,kgs+11,rkf+12,plt+13,hpm+14,ssm+14}, and there are now an appreciable number of GRB host galaxies with near-solar or super-solar metallicities \citep{pcd+07,lkg+10,srg+12,ekg+13,gf13}. Although these more massive and metal-rich GRB host galaxies do not represent the majority of the GRB host galaxy sample, they do call for certain details of the classically accepted collapsar model to be reviewed, such as the effect of differential rotation \citep{gem+12}, and binary system models \citep{pmn+04,irt04,ply+05,hp13}, or alternatively, for a re-assessment of how representative the galaxy-whole, or even the GRB line of sight properties are of the conditions within the GRB local environment.
 
From an observational perspective, sharp galaxy metallicity gradients, or low-metallicity `bubbles' within the GRB vicinity have been proposed \citep[e.g.][]{cdl+09,cts+11,kwm09}, in order to consolidate the more chemically evolved and dust-rich GRB environments measured in some cases with predications of sub-solar progenitors. However, there is thus far limited evidence to suggest that GRBs reside within metal-poor regions of their overall metal-rich galaxies \citep[although see][]{pp13}. Some spatially resolved analysis of GRB host galaxies has been done to map out the optical stellar light \citep{fls+06,hfs+06,slt+10} and dust emission \citep{mhp+14}, providing some measure of the variation in environmental properties at the GRB site relative to the rest of the host galaxy. In a few cases spatially resolved spectral analysis has been possible by either using well-positioned single slit observations \citep{hfs+06,tfo+08,lbs+11}, or in rarer cases, integral field unit (IFU) data \citep{cvs+08,tcp+14}. However, in most cases these studies have involved relatively metal-poor host galaxies. The one exception was the super-solar host galaxy of GRB~020819B\footnotemark[2], for which \citet{lkg+10} attained spectra at the GRB explosion site and the host galaxy nucleus, but found the metallicity to be the same in both regions.
\footnotetext[2]{Although this GRB was originally designated GRB~020819A, it is in fact the second GRB to have been detected that day, and is thus 020819B. GRB~020819A was detected by IPN satellites $\sim 7$~h prior to GRB~020819B. See http://www.ssl.berkeley.edu/ipn3/masterli.txt for details.}

In this paper, we report on our UV through to NIR X-shooter spectroscopic observations \citep{ddm+06,vdd+11} of the host galaxy of GRB~130925A, targeting precisely to explore further the local environments of heavily dust-extinguished GRBs. Observations began 3.5~h after the {\it Swift} GRB trigger, and although by this time the heavily dust-extinguished afterglow was barely detected, our X-shooter slit was positioned such that it covered two spatially distinct star-forming regions within the host, one of which was coincident with the GRB location.

The structure of the paper is as follows. In section~\ref{sec:obs} we describe briefly our GROND and X-shooter observations, and the X-shooter data reduction is described in section~\ref{sec:dtred}. We present our X-shooter spectral data analysis and results in section~\ref{sec:results} and in section~\ref{sec:disc} we discuss the implications that these results have on our understanding of the environments of ultra-long duration GRBs, and their progenitors. Throughout the paper, we assume a $\Omega_M=0.27$, $\Omega_\Lambda=0.73$ cosmology, with Hubble constant $H_0=70$~km~s$^{-1}$~Mpc$^{-1}$, and all reported errors are $1\sigma$.

\section{GRB and host galaxy observations}
\label{sec:obs}
GRB~130925A was detected by numerous gamma-ray detectors \citep{fitz13,hga+13,jenk13,sbf+13,ssn+13}, and the {\em Swift} burst alert telescope trigger \citep[BAT;][]{bbc+05} led to an immediate slew and subsequent X-ray afterglow detection with the X-ray telescope \citep[XRT;][]{bhn+05} at the refined position RA(J2000.0)=02:44:42.91, Dec (J2000.0) = $-$26:09:10.8 with an error radius of $1.6\arcsec$ \citep{ego+13}. No optical afterglow redward of the $R$-band was detected, with the ultraviolet and optical telescope observations \citep[UVOT;][]{rkm+05} taken just 157s after the GRB trigger, resulting in a $v$-band $3\sigma$ upper limit of $v>20.0$ \citep{hl13}. Rapid-response observations with the simultaneous 7 channel imager, GROND, began just 7~minutes after the GRB trigger \citep{skg13}, and a decaying optical source was detected within the X-ray afterglow error circle in the $i',z',J,H$ and $K$ bands \citep{skg13,gyk+14}. A detailed analysis of the XRT and GROND afterglow observations are given in \citet{ewo+14} and \citet{gyk+14}, respectively.

Rapid response mode spectroscopic observations of GRB~130925A with the VLT/UVES spectrograph began 50~minutes after the {\em Swift} trigger, and revealed faint continuum emission and two emission lines identified as H$\alpha$ and [N~{\sc ii}] at a redshift of $z=0.347$ \citep{vmf+13}. Owing to the lack of a bright afterglow, and thus of absorption lines, the UVES observations were aborted, enabling us to trigger our X-shooter ToO programme (ID 091.A-0703; PI: P. Schady).

Our X-shooter spectroscopic observations began on 25-09-2013 at 07:43~UT, 3.5~h after the BAT trigger. Because of the disabling of the X-shooter atmospheric dispersion correctors in August 2012, the slit position was orientated to the parallactic angle in order to minimise atmospheric dispersion losses. The observations consisted of four nodded exposures in the sequence ABBA with exposure times of $\sim 1500$s taken simultaneously in X-shooter's untraviolet/blue (UVB), visible (VIS) and near-infrared (NIR) arms. The average airmass during our observations was 1.0 and the seeing was a median of 0.6\arcsec. X-shooter spectroscopy was performed with slit-widths of 1.\arcsec0, 0.\arcsec9 and 0.\arcsec9 in the UVB, VIS and NIR arm, respectively. The resolving power $R=\lambda/\Delta\lambda$ was determined from arc lamp calibration data taken with the same instrumental setup as our science observations\footnotemark[3]. From this we get $R\sim 5300$ for the VIS and NIR arms, and $R\sim 8800$ for the VIS arm. These correspond to an instrumental velocity FWHM of $\sim 55$~km~s$^{-1}$ and $\sim 35$~km~s$^{-1}$, respectively.
\footnotetext[3]{archive.eso.org/bin/qc1\_cgi}

\begin{figure}
\centering
\includegraphics[width=0.5\textwidth]{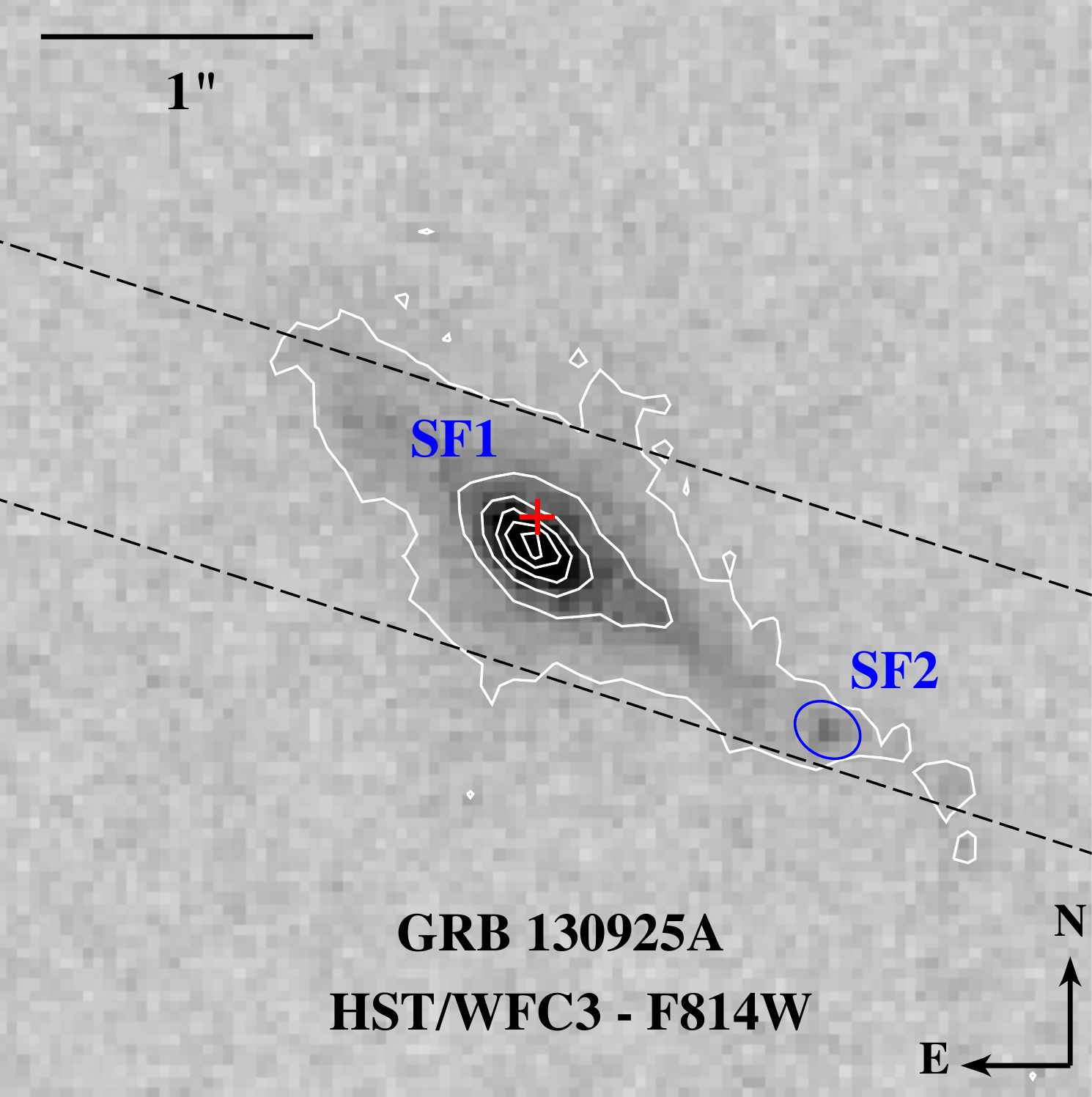}
\caption{HST/F814W image of the host galaxy of GRB~130925A. The image is 4\arcsec$\times $4\arcsec centred on the GRB host galaxy. It is displayed with a linear greyscale ranging from $0$~count~s$^{-1}$~pix$^{-1}$ (white)  to $0.1$~count~s$^{-1}$~pix$^{-1}$ (black), and logarithmically spaced contours are shown in white. The position of the X-shooter's VIS slit with width 0.\arcsec9 is indicated by the black dashed lines. The UVB and NIR slits have the same orientation. The position of the afterglow was determined from image subtraction between two HST/F814W observations taken 20 and 48~days after the GRB, and is indicated by the red cross. It falls within the emission region SF1 detected in our spectral data, which in the above image, corresponds roughly to the second dimmest contour plotted. The emission region, SF2, detected in our spectra and located 1.\arcsec2 from SF1 is indicated at the SW corner of the outer contour plot by a blue ellipse.}\label{fig:fc}
\end{figure}

HST observations taken 20 days after the GRB indicate that the GRB lies close to the plane of the disk of what appears to be an edge-on spiral, just $\sim 0.\arcsec 12$ from the nucleus ($\sim 600$~pc in projection) \citep{tlh+13}. A detailed discussion on the HST observations of the host galaxy will be reported in Tanvir et al. (in prep.). We also observed the field of GRB~130925A with the VLT/HAWK-I instrument in the $Y$, $J$, $H$ and $K_S$ bands 18 days after the GRB trigger (ID 092.A-0231; PI: T. Kr{\"u}hler), and the measured photometry was consistent with the source magnitudes measured seven days earlier with GROND, suggesting that these late-time GROND and HAWK-I observations were dominated by emission from the host galaxy. From our more sensitive HAWK-I data we measure host galaxy AB smagnitudes of $J_{AB}= 20.70\pm 0.07$~mag, $H_{AB} = 20.47\pm 0.07$~mag and $K_{AB} = 20.20\pm 0.08$~mag \citep{gyk+14}, corresponding to an absolute (K-corrected) $H$-band magnitude of $m_H(AB)=-20.6\pm 0.08$~mag. The HST F814W image with the position of the X-shooter slit overlaid is shown in Fig.~\ref{fig:fc}.

%%%%%%%%%%%%%%%%%%%%%%%%%%%%%%%%%%%%%%%%%%%%
%% FIGURE 1
%%%%%%%%%%%%%%%%%%%%%%%%%%%%%%%%%%%%%%%%%%%%

\section{X-Shooter data reduction}
\label{sec:dtred}
X-shooter data were reduced with the ESO/X-shooter pipeline v2.2.0 \citep{grf+06}, including the flat-fielding, order tracing, rectification and initial wavelength calibration from arc lamp data. During rectification, a dispersion of 0.2~\AA/pix and 0.6~\AA/pix was used in the UVB, VIS and NIR arms, respectively, which minimises correlated noise while still maintaining sufficient spectral resolution to resolve lines down to $\sim 25$~km~s$^{-1}$ in the UVB and VIS arms (i.e. a velocity dispersion of 10~km~s$^{-1}$), and down to $\sim 50$~km~s$^{-1}$ in the NIR arm (i.e. a velocity dispersion of 20~km~s$^{-1}$). Our own software was used for bad pixel and cosmic-ray rejection, for sky subtraction, and for frame shifting and adding (see \citealt{kmf+15} for details). The X-shooter spectra were then flux calibrated with the nightly spectrophotometric standard star and optimally extracted within IRAF using a large aperture in order to include all the emission from the galaxy. The 1$\sigma$ error bars on each data point in our extracted 1D spectra were derived via the error spectrum and propagated accordingly.

%%%%%%%%%%%%%%%%%%%%%%%%%%%%%%%%%%%%%%%%%%%%
%% FIGURE 2
\begin{figure*}
\centering
\includegraphics[width=0.36\textwidth]{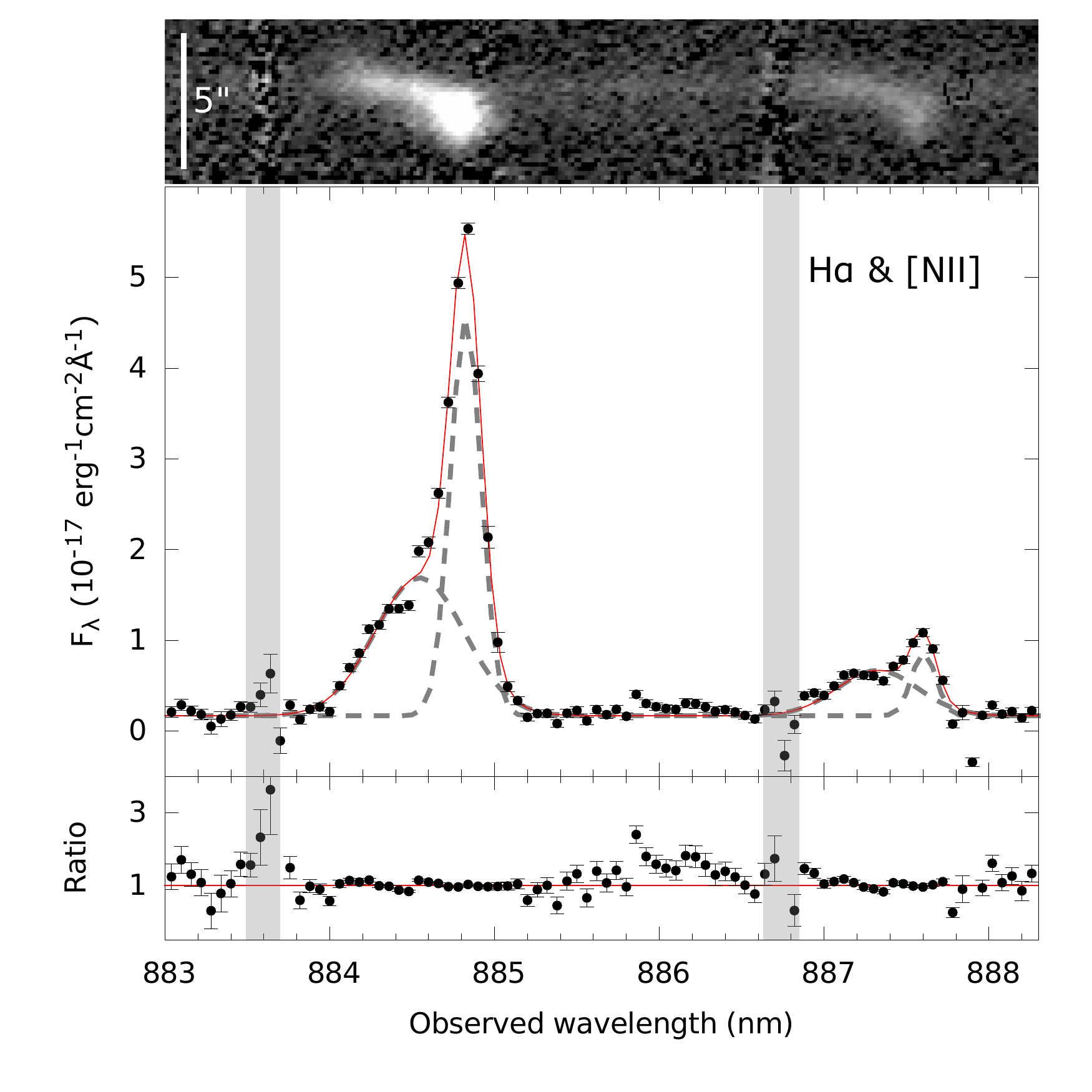}
\includegraphics[width=0.27\textwidth]{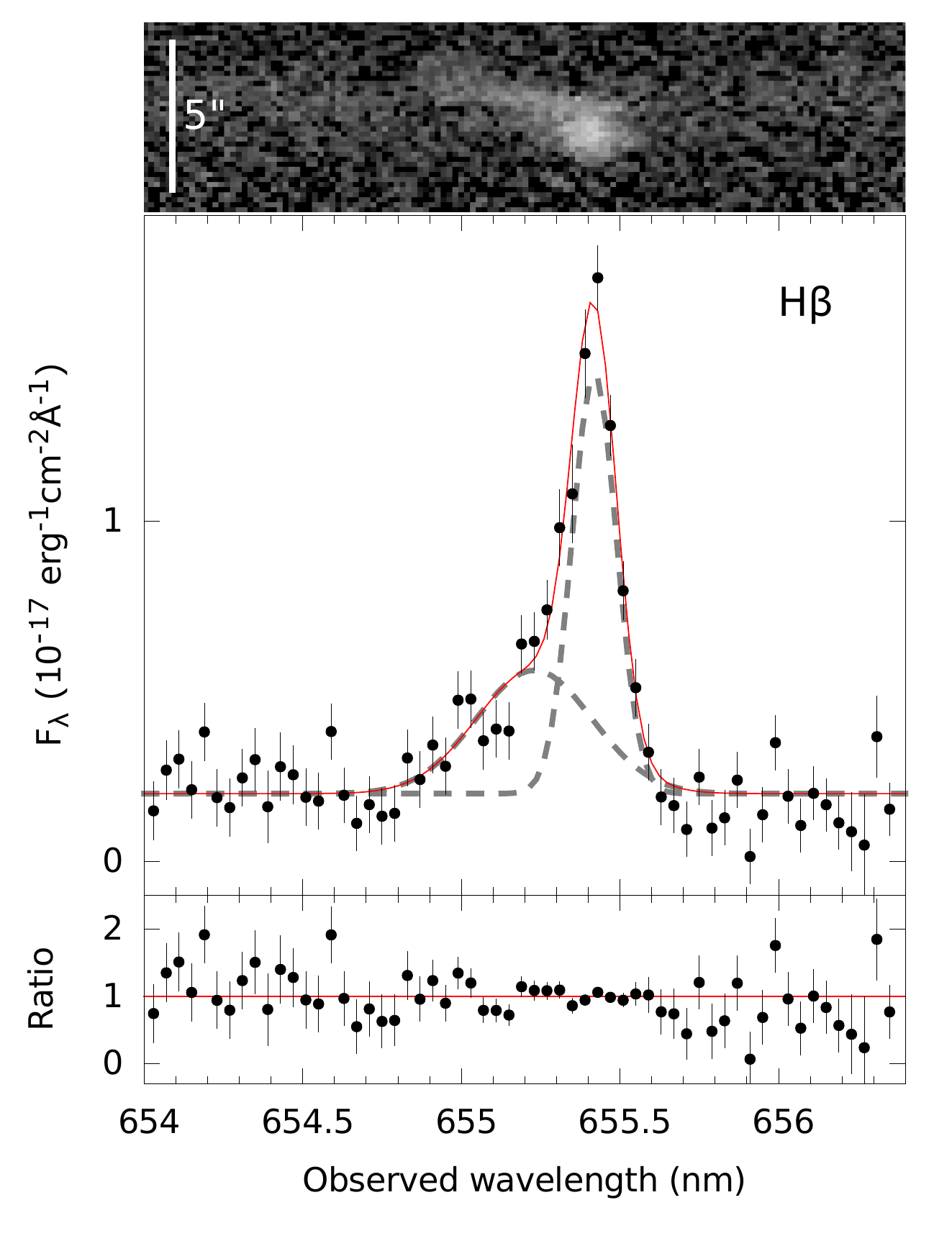}
\includegraphics[width=0.27\textwidth]{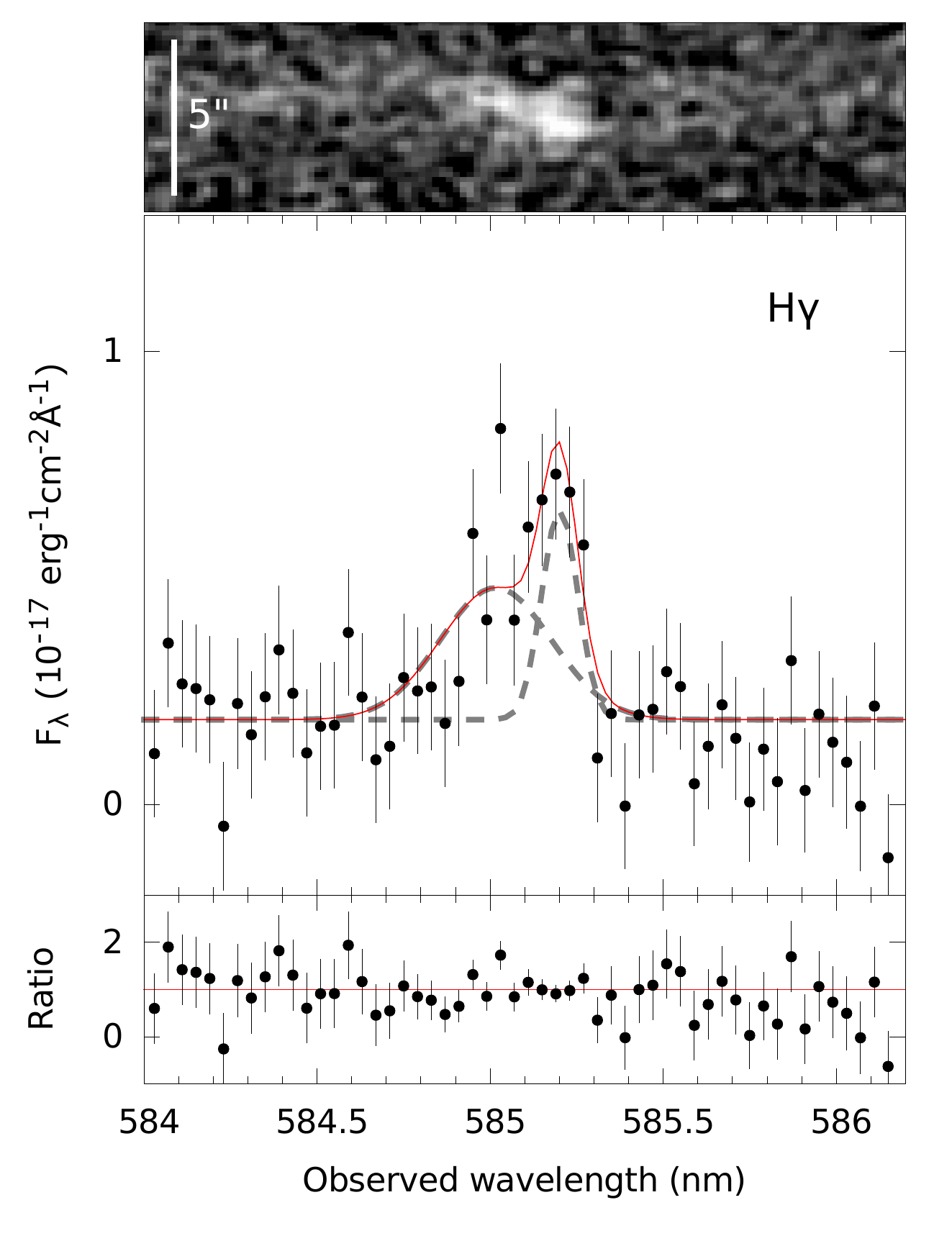}
\includegraphics[width=0.27\textwidth]{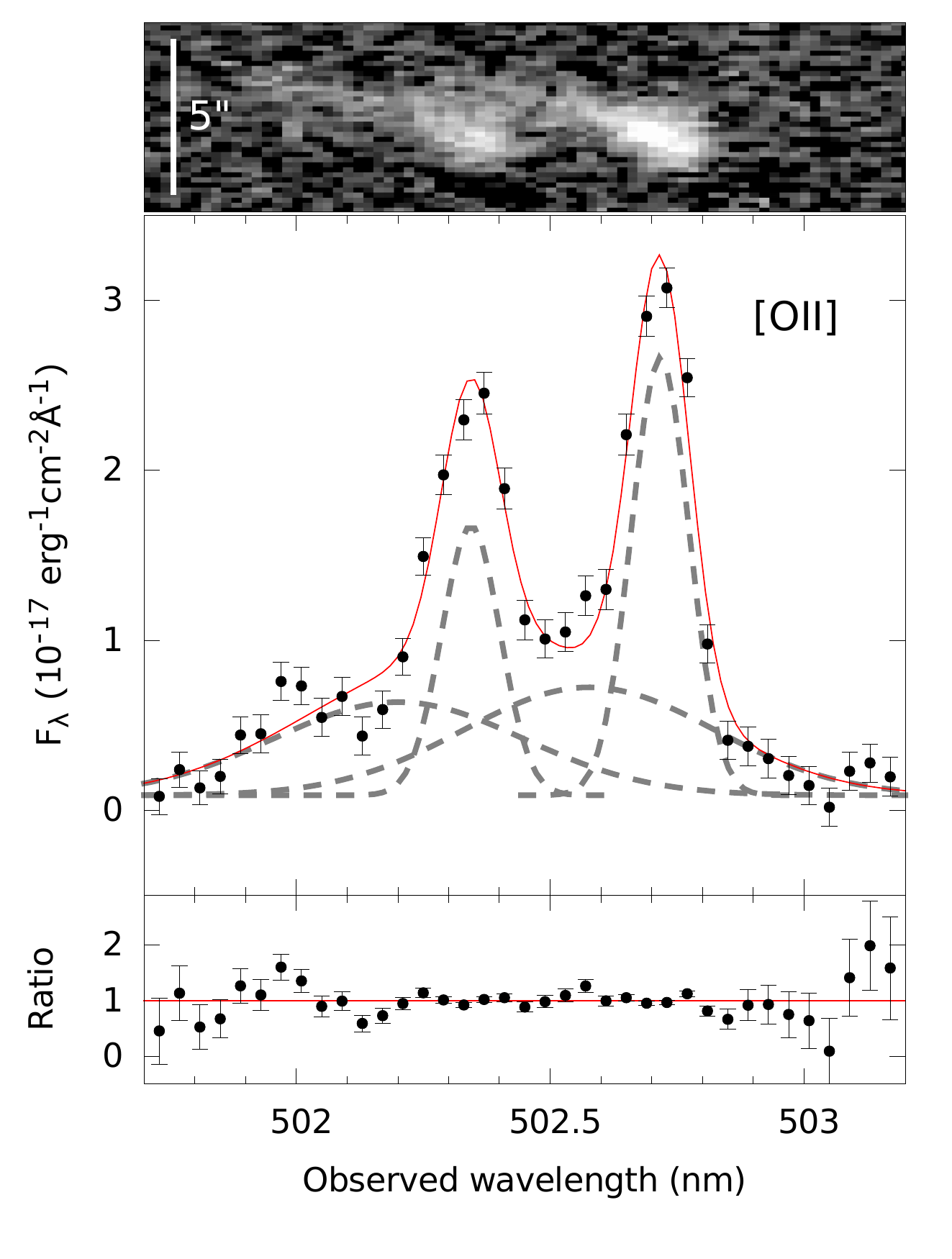}
\includegraphics[width=0.36\textwidth]{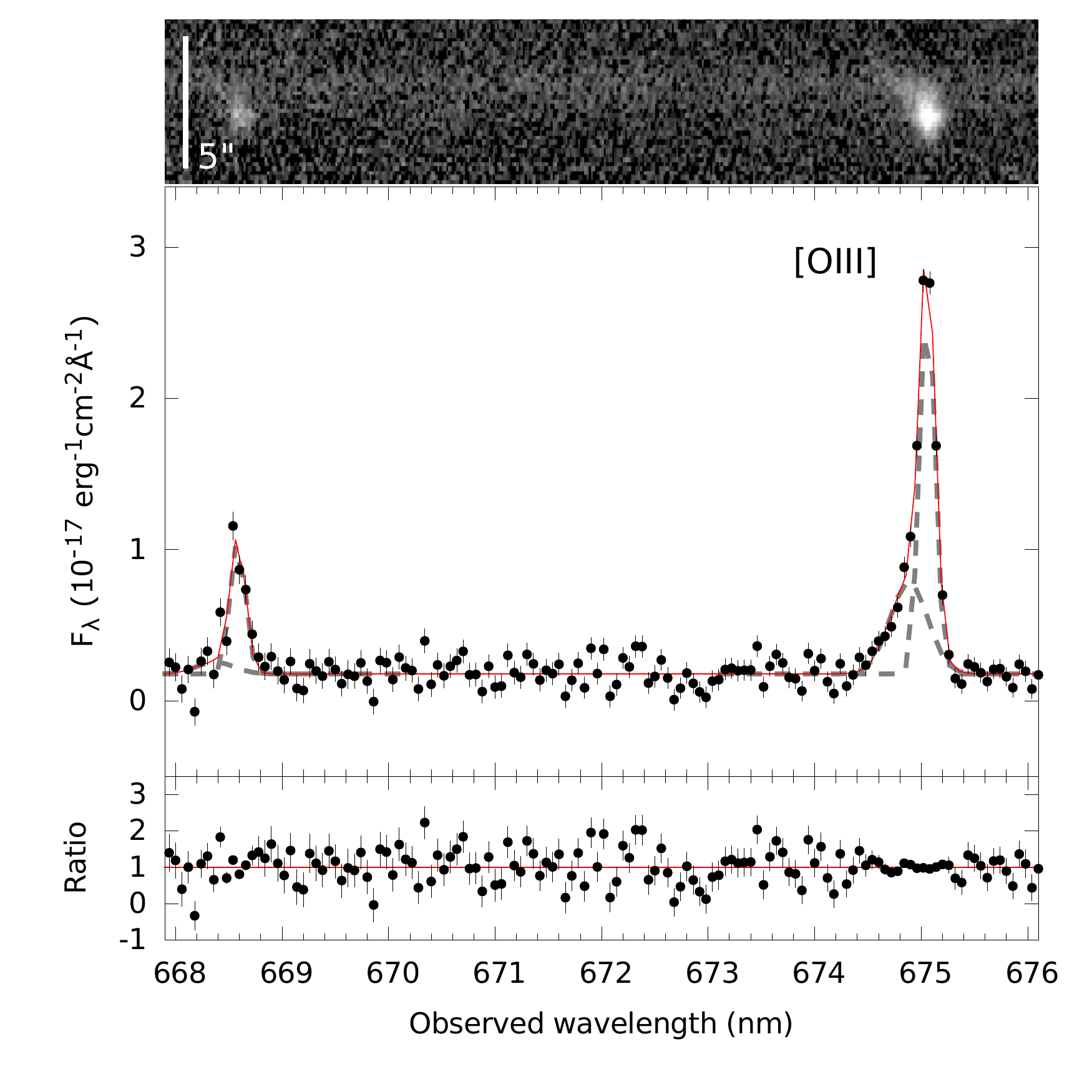}
\includegraphics[width=0.36\textwidth]{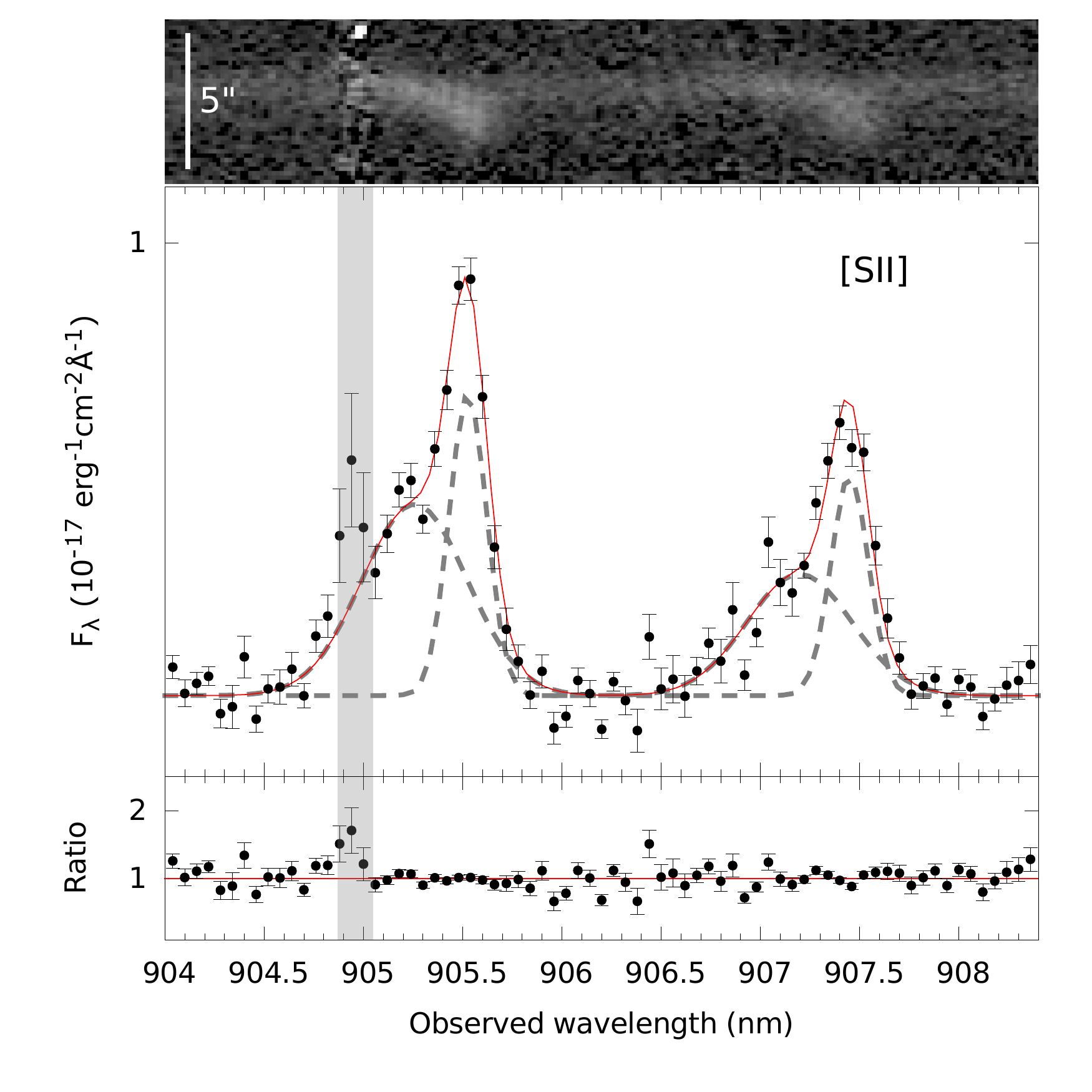}
\caption{Balmer and forbidden emission lines from the host galaxy of GRB~130925A. In each figure, the top and middle panels show the X-shooter 2D and 1D spectra zoomed into the wavelength range of the line labelled in the top right corner of the middle panel. The bottom panel shows the residuals from our two-component fits to the 1D spectrum. In the middle panel, we also show the best-fit Gaussians for each individual component (dashed grey lines) and composite model (red solid line).}\label{fig:lines}
\end{figure*}
%%%%%%%%%%%%%%%%%%%%%%%%%%%%%%%%%%%%%%%%%%%%

\section{Data analysis and results}
\label{sec:results}
The X-shooter spectrum of GRB~130925A is dominated by emission lines from the host galaxy, with only a very weak continuum detected, corresponding mostly to emission from the host galaxy, with a small contribution in the NIR arm from the afterglow. This is verified by the continual GROND observations taken between 350s and $10^6$s after the GRB trigger, which show that at the epoch of our X-shooter observations, 3.5~h after the GRB trigger, the emission predominately originates from the host galaxy light, with the afterglow contributing only a small amount to the flux in the GROND $J$, $H$ and $K$ bands \citep{gyk+14}. In both the 2D and the 1D spectra, we clearly detect all the usual strong emission lines, and H$\beta$, H$\gamma$, [N~{\sc ii}] and the [S~{\sc ii}]($\lambda$6717 \& $\lambda$6731) doublet are all detected at better than 3$\sigma$ (see Fig.~\ref{fig:lines}). In our 2D spectra, emission from at least two star-forming (SF) regions, SF1 and SF2, is detected, corresponding to emission from the central disk/bulge of the galaxy, and from an HII region located $\sim 1.$\arcsec2 south-west of the galaxy nucleus (i.e. 6~kpc in projection; see Fig.~\ref{fig:fc}). The GRB afterglow position is coincident with the location of the weaker and extended SF1 region.

We extracted a single 1D spectrum for each of the X-shooter arms using a large aperture of width $4.\arcsec2$, which contains the emission from both SF1 and SF2. The emission profiles for most of the detected lines clearly deviate from a Gaussian, and we therefore fitted each emission line in our 1D spectra with two Gaussians simultaneously, one per detected emission region. For line doublets we tied the Gaussian width of the lines corresponding to the same emission component, and in the case of the weak H$\gamma$ line, we fixed the Gaussian width and central line peak positions for the two emission components to correspond to the best fit redshift and velocity line width determined from our fits to the H$\alpha$ line. In our fits to the [O~{\sc ii}] and [S~{\sc ii}] doublets, and to the [O~{\sc iii}]$\lambda$4959 line, we fixed the central line peak position of the weaker Gaussian component from SF1 to correspond to the best fit redshift measured at SF1 in our H$\alpha$ line fits.

Slit losses on our line emission flux measurements due to the finite width of the slit were estimated independently for each arm against the multi-band GROND host galaxy photometry, and were found to be consistent within errors. The corresponding average slit loss correction was a multiplicative factor of $2.7\pm 0.1$. The spatial offset between the two detected emission regions introduces further uncertainty in our slit-loss correction, which is likely to not be the same for the two star-forming regions. Given the proximity of SF2 to the slit edge (see Fig.~\ref{fig:fc}), the respective slit losses might be notably larger than that at SF1. This will not affect greatly the dust extinction or metallicity that we measure at SF1 and SF2, which is the focus of this paper. However it may impact our measured line fluxes (see Table~\ref{tab:lineflux}) and derived star-formation rate (SFR), such that they are over-estimated at SF1 and under-estimated at SF2.

We are unable to resolve and thus correct our measured Balmer line fluxes for Balmer absorption \citep[e.g.][]{sgl+05,cmi+06}. However, for such a weak stellar continuum, we expect the effect of Balmer absorption on our line fluxes to be small, with other factors such as slit loss and dust extinction corrections being a more dominant source of uncertainty. Our best-fit slit loss corrected line fluxes are listed in Table~\ref{tab:lineflux}, where the given errors include the uncertainties on the location of the Gaussian peak flux and the FWHM. Systematic photometric calibration errors and slit-loss correction errors are not included, which correspond to a further 5-8\%.

%%%%%%%%%%%%%%%%%%%%%%%%%%%%%%%%%%%%%%%%%%%%
%% TABLE 1
\begin{table}
\begin{center}
\caption{GRB~130925A host galaxy line fluxes at SF1 and SF2 corrected for a Galactic reddening of E(B-V)=0.02\label{tab:lineflux}}
\begin{tabular}{@{}lccc}
\hline
 & Wavelength$^a$ & $F_\lambda($SF1$)$ & $F_\lambda($SF2$)$ \\
 & \AA & \multicolumn{2}{c}{($10^{-17}$~erg~cm$^{-2}$~s$^{-1}$)}\\ 
\hline\hline
H$\gamma$ & 4340.5 & $3.40^{+0.26}_{-0.26}$ & $1.77^{+0.19}_{-0.19}$ \\
H$\beta$ & 4861.3 & $4.81^{+0.50}_{-0.44}$ & $6.30^{+0.11}_{-0.11}$ \\
H$\alpha$ & 6562.8 & $27.80^{+1.56}_{-1.60}$ & $27.67^{+0.10}_{-0.10}$ \\
 $[$O~{\sc ii}$]$ & 3726.0 & $10.83^{+1.31}_{-1.11}$ & $7.29^{+0.11}_{-0.10}$ \\
 $[$O~{\sc ii}$]$ & 3728.8 & $12.55^{+1.49}_{-1.26}$ & $11.84^{+0.11}_{-0.11}$ \\
 $[$O~{\sc iii}$]$ & 4958.9 & $1.01^{+0.18}_{-0.19}$ & $4.57^{+0.12}_{-0.11}$ \\
 $[$O~{\sc iii}$]$ & 5006.8 & $7.81^{+0.11}_{-0.11}$ & $12.96^{+0.10}_{-0.10}$ \\
 $[$N~{\sc ii}$]$ & 6583.4 & $5.61^{+0.30}_{-0.47}$ & $3.46^{+0.11}_{-0.11}$ \\
 $[$S~{\sc ii}$]$ & 6716.5 & $6.06^{+0.76}_{-0.69}$ & $3.42^{+0.05}_{-0.05}$ \\
 $[$S~{\sc ii}$]$ & 6730.9 & $3.85^{+0.49}_{-0.45}$ & $2.50^{+0.05}_{-0.05}$ \\
\hline
\multicolumn{3}{l}{$^a$ Rest-frame vacuum wavelength}
\end{tabular}
\end{center}
\end{table}
%%%%%%%%%%%%%%%%%%%%%%%%%%%%%%%%%%%%%%%%%%%%

\subsection{Velocity dispersion and dynamical mass}
\label{ssec:sigma}
Our two-component spectral fits give best-fit Gaussian widths corresponding to velocity dispersions of $80-90$~km~s$^{-1}$ at SF1, and $25-30$~km~s$^{-1}$ at SF2, which is close to the velocity resolution of our data. Such velocities are at the lower and higher $\sim 20$\% of the distribution of velocity dispersions observed in GRB host galaxies at redshifts $z<1$ \citep{kmf+15}.
From our best-fit emission line central wavelengths we get heliocentric\footnotemark[4] redshifts of $z_{SF1}=0.3479$ and $z_{SF2}=0.3483$ at SF1 and SF2, respectively, corresponding to a separation in velocity space of 90~km~s$^{-1}$. Within the projected distance of $\sim 6$~kpc of SF2 from the galaxy nucleus, a velocity dispersion of 90~km~s$^{-1}$ corresponds to a logarithmic dynamical mass of $\log [M_{dyn}/M_\odot]\sim10.1$.
\footnotetext[4]{The heliocentric correction in the direction of GRB~130925A is 10~km~s$^{-1}$ for our observations.}

\subsection{Balmer decrement and SFR}
\label{ssec:balmdec}
Assuming a temperature T=$10^4$~K and a density of $10^2-10^4$~cm$^{-3}$, we compare the observed H$\alpha/$H$\beta$ Balmer decrement to the intrinsic, unextinguished ratio of 2.86 in order to measure the host galaxy visual extinction, A$_V$, at SF1 and SF2. Assuming a Calzetti dust attenuation law \citep{cab+00}, we measure A$_V$=$2.4\pm 0.9$~mag and A$_V$=$1.5\pm 0.1$~mag at SF1 and SF2, respectively. Comparing the measured $H\alpha$/H$\gamma$ ratio to the intrinsic ratio of 6.11 results in visual extinctions of A$_V$=$0.7\pm 0.2$~mag and $2.3\pm 0.6$~mag at SF1 and SF2, respectively, which are consistent at $2\sigma$ with the A$_V$ values determined from the measured H$\alpha$/H$\beta$ ratio. Given the relative weakness of the H$\gamma$ emission line, and thus the larger uncertainty in the corresponding flux density measurement, we choose to use the A$_V$ values derived from the H$\alpha/$H$\beta$ ratio for SF1 and SF2. This gives us a dust-corrected H$\alpha$ luminosity of $(7.1\pm 5.2)\times 10^{41}$~erg~s$^{-1}$ and $(3.6\pm 0.3)\times 10^{41}$~erg~s$^{-1}$ at SF1 and SF2, respectively, corresponding to SFRs of $3.4\pm 2.5$~M$_\odot$~yr$^{-1}$ and $1.7\pm 0.1$~M$_\odot$~yr$^{-1}$ based on the formalism described in \citet{ken98}, but converted to a Chabrier IMF \citep{cha03}, which reduces the SFR by a factor of $\sim 1.7$ \citep[e.g.][]{fgb+09}.

\subsection{Host galaxy metallicity}
\label{ssec:metal}
There are a number of available empirical and theoretical metallicity diagnostics, each with their own strengths and drawbacks. Owing to the range in emission line ratios and physical assumptions (e.g. ionisation parameter) that go into the different gas-phase metallicity measurements, certain diagnostics are more appropriate for sources within a given metallicity or redshift interval. Nevertheless, which of the metallicity diagnostics is used is typically limited by the emission lines detected for the source in question, and thus it is not always possible to use the most accurate metallicity diagnostic. As a result of the very large wavelength range available with X-shooter, we measure all the relevant emission lines (Table~\ref{tab:lineflux}) required from the two star forming regions to apply several of the more accurate metallicity diagnostics.

In a detailed investigation on the accuracy and cross-calibration between the more popular theoretical and empirical metallicity diagnostics, \citet{ke08} derive metallicity conversions between eight of the strong-line metallicity calibrations, allowing the metallicities to be converted to the same base calibration. Nevertheless, such conversions are based on the average difference between metallicity diagnostics, and are therefore designed to be used on galaxy samples, rather than on individual sources. Below we therefore derive the host galaxy metallicity at SF1 and SF2 for a number of diagnostics, and give the metallicities using the native metallicity diagnostic calibration, as well as the metallicities converted to the \citet{kk04} scale, applying the conversions from \citet{ke08}. The latter conversion to a base scale facilitates a comparison between the derived metallicities and solar metallicity, which we take to be $\log($O/H$)+12 = 8.69\pm 0.05$ \citep{ala+01,ags+09}. The $R_{23}$ diagnostic from \citet{kk04} is calibrated against this same solar metallicity (see their footnote 2), which is why we choose this metallicity diagnostic as our base scale.

\citet{ke08} found that the N2O2 metallicity diagnostic as calibrated by \citet{kd02} was the most reliable for systems with $\log [$N~{\sc ii}/O~{\sc ii}$]>-1.2$ (corresponding to $12+\log [$O/H$]>8.4$) because of its weak dependence on ionisation parameter within this metallicity regime. For $12+\log [$O/H$]<8.4$ the dependence on ionisation parameter increases, and thus in such a case \citet{ke08} recommend using the $R_{23}$ method, which estimates the ionisation parameter from the observed $[$O~{\sc iii}$]/[$O~{\sc ii}$]$ line ratio. The drawback of the $R_{23}$ method is that it has a lower and an upper branch solution. However, the degeneracy between the two solutions can be broken using other metallicity diagnostics, such as in this case by the N2O2 diagnostic, whereby the lower branch solution is appropriate for sources with a flux line ratio $\log [$N~{\sc ii}/O~{\sc ii}$]<-1.2$.

In the case of the host galaxy of GRB~130925A, we measure $\log [$N~{\sc ii}/O~{\sc ii}$]=-0.7$ at SF1 and SF2, after correcting for host galaxy dust reddening. The \citet{kd02} N2O2 diagnostic is thus appropriate, and gives us a metallicity of $12+\log [$O/H$]_{\rm N2O2}=8.77^{+0.04}_{-0.02}$ at SF1, and $12+\log [$O/H$]_{\rm N2O2}=8.79^{+0.01}_{-0.01}$ at SF2. In the \citet{kk04} scale, this corresponds to  $12+\log [$O/H$]_{\rm N2O2}=8.86^{+0.04}_{-0.02}$ and $12+\log [$O/H$]_{\rm N2O2}=8.88^{+0.01}_{-0.01}$ at SF1 and SF2 respectively \citep{ke08}, which is equivalent to $1.5Z_\odot$. The reported errors include the uncertainties in the line flux measurements and host galaxy reddening correction derived from the Balmer decrement, but not the systematic calibration error of $\sim 0.1$~dex.

\citet{ke08} also found the O3N2 and N2 diagnostics from \citet{pp04} to be fairly robust, with the added advantage that these methods are fairly insensitive to the dust reddening correction applied as a result of the proximity in wavelength space of the lines used. This feature makes these two diagnostics favourable for high redshift ($z>1$) sources, where there may be a large uncertainty in the measured Balmer decrement. Although this is not a significant issue in our data, we nevertheless calculate the metallicity measured with these two diagnostics, and find that $12+\log [$O/H$]_{\rm O3N2}=8.50\pm 0.03$ and $12+\log [$O/H$]_{\rm N2}=8.60\pm 0.04$ at SF1, and $12+\log [$O/H$]_{\rm O3N2}=8.35\pm 0.01$ and $12+\log [$O/H$]_{\rm N2}=8.33\pm 0.01$ at SF2. In the \citet{kk04} scale, the metallicity at SF1 is $12+\log [$O/H$]_{\rm O3N2}=8.85\pm 0.03$ and $12+\log [$O/H$]_{\rm N2}=8.90\pm 0.04$, and it is $12+\log [$O/H$]_{\rm O3N2}=8.73\pm 0.01$ and $12+\log [$O/H$]_{\rm N2}=8.72\pm 0.01$ at SF2. These values correspond to respective O3N2 and N2 metallicities of $1.4Z_\odot$ and $1.6Z_\odot$ at SF1, and to $1.1Z_\odot$ at SF2 in both the O3N2 and N2 metallicity diagnostics. When considering the respective systematic calibration errors of $\sim 0.2$~dex and $\sim 0.4$~dex for the O3N2 and N2 methods, the derived metallicities are consistent at $1\sigma$ with those derived with the N2O2 calibrator.

Given that the $R_{23}$ method is frequently used, for completeness we also use this diagnostic to measure the metallicity in SF1 and SF2, applying the calibration from \citet{kk04}, which simultaneously fits the ionisation parameter and gas-phase metallicity. This gives us an upper-branch solution of $12+\log [$O/H$]_{R_{23}}=8.70^{+0.12}_{-0.16}$ ($\sim Z_\odot$) at SF1, and $12+\log [$O/H$]_{R_{23}}=8.90\pm 0.01$ ($1.6Z_\odot$) at SF2, which is fully consistent with the N2O2, O3N2 and N2 derived metallicities when systematic errors are taken into account. 

A solar or above metallicity is also supported by the lack of emission detected at the location of the [O~{\sc iii}]$\lambda 4363$ emission line down to a $3\sigma$ upper limit of $<1.4e^{-17}$~erg~cm$^{-2}$~s, whereas we would expect to detect emission from this line within a low-metallicity environment.

\subsection{Galaxy spectral energy distribution}
\label{ssec:sed}
\begin{figure}
\centering
\includegraphics[width=0.5\textwidth]{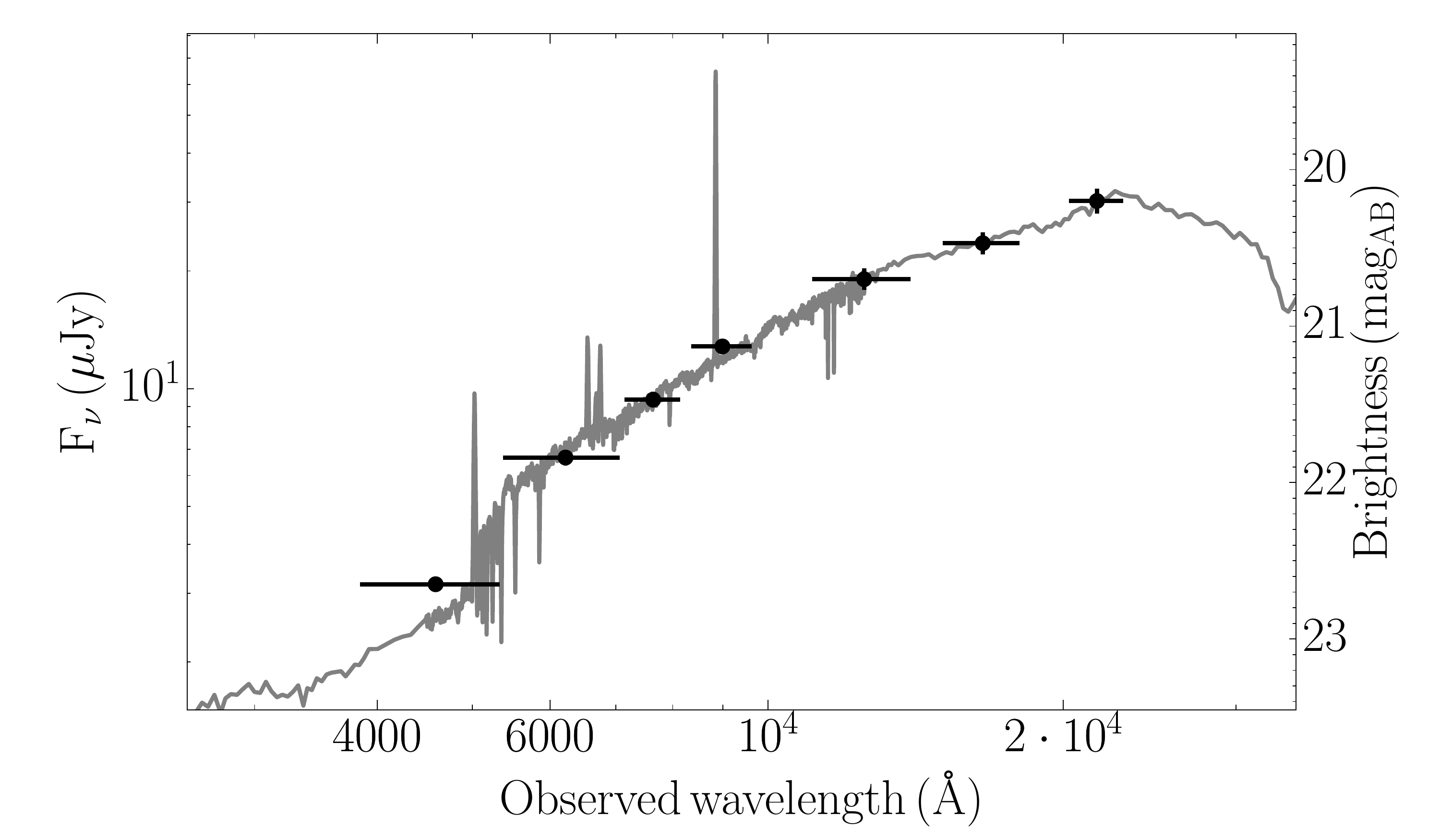}
\caption{SED of the host galaxy of GRB~130925A and best-fit galaxy model (solid line) in the observer frame. Filled black circles represent photometric measurements taken with GROND in $g'r'i'z'$ and HAWK-I in $JHK$.}\label{fig:hostsed}
\end{figure}

Using late time $g'r'i'z'JHK$ GROND and $JHK$ VLT/HAWK-I data taken 11~days and 18~days after the GRB trigger, respectively \citep{gyk+14}, we created the host galaxy SED. There was no evidence of further fading in the $JHK$ filters between these two epochs, indicating that we were detecting the host galaxy at these times with a negligible contribution from the afterglow. The flux in the optical bands plateaued at 2~days after the GRB trigger, with GROND observations taken 2, 4 and 11~days after the trigger yielding consistent magnitudes \citep{gyk+14}. We modelled the host galaxy SED using the spectral fitting package {\em LePHARE} \citep{acm+99,iam+06}, which is a population-synthesis-based fitting procedure. We used the \citet{bc03} galaxy templates, which include emission lines and prescribed reddening and parameters therein, and we assumed a \citet{cha03} IMF (see Fig.~\ref{fig:hostsed}). We measure a host galaxy logarithmic stellar mass of $\log [M_\ast/M_\odot]=9.5\pm 0.2$ and SFR of $2.4^{+1.6}_{-1.3}$~M$_\odot$~yr$^{-1}$ for a host galaxy reddening of $E(B-V)=0.2$, assuming again a Calzetti dust attenuation law \citep{cab+00}. This is consistent within the errors with the H$\alpha$ luminosity-derived SFR (section~\ref{ssec:balmdec}).

\subsection{Combined versus individual spectral analysis}
\label{ssec:1Dfits}
We re-did the above analysis on the individual SF1 and SF2 regions by using smaller apertures to extract the 1D spectra, centred on the peak flux of each emission region. The apertures we used were $1.\arcsec4$ and $2.\arcsec1$ wide for the SF1 and SF2 regions, respectively. From the line fluxes measured from the separate 1D spectra, we derived Balmer decrement corrections and metallicities at SF1 and SF2 fully consistent with the values measured from our combined spectral analysis, although the error bars corresponding to the metallicity at SF1 increased by a factor of around three.

%%%%%%%%%%%%%%%%%%%%%%%%%%%%%%%%%%%%%%%%%%%%
%% FIGURE 3
%%%%%%%%%%%%%%%%%%%%%%%%%%%%%%%%%%%%%%%%%%%%

\section{Discussion}
\label{sec:disc}
\subsection{Dust and metal abundances in GRB host galaxies}
The host galaxy of GRB~130925A is remarkable in terms of the amount of dust attenuation detected along the disk of the galaxy (A$_V$$= 2.4\pm 0.9$~mag), as well as in the outer regions of the galaxy (A$_V$$=1.5\pm 0.1$~mag). The host galaxy dust-extinction measured along the GRB line of sight from the afterglow SED (A$_V$=$5.0\pm 0.7$~mag) is the second largest thus far measured in a GRB afterglow\footnotemark[5], not considering A$_V$ lower limits.
\footnotetext[5]{The afterglow of GRB~070306 currently has the largest measured host galaxy, afterglow visual extinction, with A$_V$$=5.5^{+1.2}_{-1.0}$~mag \citep{jrw+08,kgs+11}.}
The relatively low redshift of this GRB makes the high afterglow extinction all the more unusual, given that more heavily dust-extinguished GRB afterglows, and correspondingly dust-rich GRB host galaxies, are more numerous at high redshift \citep{plt+13,kmf+15}. Whereas at $z<1$ around 20\% of GRB host galaxies have a visual extinction $A_V>1.2$~mag, and less than 10\% have $A_V>1.5$~mag, at redshifts $z=1-2$, $\sim 60$\% of GRB host galaxies have $A_V>1.2$~mag, and over 40\% have $A_V>1.5$~mag \citep{kmf+15}. Prior to the commissioning of NIR, rapid-response facilities, such as GROND and PAIRITEL, the afterglow of this GRB at wavelengths longward of X-rays would likely not have been detected.

The proximity of the GRB to the nucleus of the galaxy, and more importantly, the seemingly edge-on orientation of the host galaxy implies that the GRB line-of-sight crossed a particularly dusty region of its host, which may, at least in part, account for the unusually high GRB afterglow visual extinction. Nevertheless, it is noteworthy that GRB~130925A is only the fourth GRB host galaxy at $z<0.5$ to have solar or above metallicity, and that a high metallicity is measured within the disk, and in a star-forming region in the outskirts of the galaxy. These observations make the host galaxy of GRB~130925A only the second super-solar host galaxy where spatially resolved spectroscopic observations have been possible; the first being the host galaxy of GRB~020819B at $z=0.411$ \citep{lkg+10}. Furthermore, unlike the spectroscopic observations of the host galaxy of GRB~020819B for example, the unique broad wavelength range of X-shooter covers all emission lines necessary to be able to measure the metallicity relatively accurately. Acquiring such measurements for metal- and dust-rich GRB host galaxies is challenging as a result of the initial difficulty in detecting the afterglow along such highly dust extinguished lines of sight, and thus acquiring an arcsecond GRB position, and secondly, in then obtaining sufficiently high signal-to-noise host galaxy spectra to measure the Balmer decrement and galaxy metallicity with any degree of accuracy.

There are several similarities in the properties of the host galaxies of GRB~020819B and GRB~130925A, including a comparably large visual dust extinction at the galaxy nucleus of A$_V$$=1.5-2.0$~mag as derived from the measured Balmer decrement, and a super-solar metallicity at the nucleus and outskirts of the galaxy. The stellar mass, SFR and metallicity of the host galaxy of GRB~020819B were, however, somewhat larger than that of host galaxy of GRB~130925A \citep[i.e. $\log M_\ast/M_\odot\sim 10.5$, SFR$\sim 15$M$_\odot$~yr$^{-1}$, $\log(O/H)+12\sim 9.0$ in the \citealt{kd02} scale;][]{sgl09,kgk+10,lkg+10,slt+10}. An important observational distinction between the two galaxies is that the host galaxy of GRB~020819B is face-on, and the GRB occurred within the outskirts of the galaxy. Spectroscopic observations of this latter host galaxy thus probe smaller radial length scales than is the case in our observations of the host galaxy of GRB~130925A. The uncertainty on how representative the spectroscopic measurements are of the conditions nearby to the GRB are thus larger in the case of GRB~130925A \citep{nnz15}. Our ability to quantify the probability that a GRB occurred in a low-metallicity region of its globally metal-rich host galaxy will improve with larger sample sizes of nearby, high-metallicity, and preferably face-on host galaxies. However, for now, the observations of the face-on host galaxy of GRB~020819B provide some support to the notion that GRBs can arise within metal-rich environments.

The increase in the detection rate of more dust and metal-rich GRB sightlines and host galaxies have already placed challenges on the standard collapsar model \citep[e.g.][]{lkg+10,kgs+11,srg+12,plt+13}. However, in addition to the spectroscopically resolved observations presented here, this host galaxy is distinctive in that it is the first example of a super-solar metallicity galaxy to host a GRB that makes up part of the currently small sample of ultra-long GRBs. As is the case for long-duration GRBs, the favoured theoretical models for the ultra-long class of GRBs place an upper limit on the progenitor metallicity.

\subsection{Ultra-long GRBs; a distinct class of GRB?}
\label{ssec:ulGRBs}
The prompt emission of GRB~130925A continued for an extraordinarily long time \citep[$\sim 20$~ks;][]{ewo+14,ptg+14}, and the X-ray afterglow light curve exhibited numerous large flares that were initially believed to originate from a TDE \citep{bml+13}. The X-ray spectrum was also unusually soft ($\beta_X\sim 3$) and an emission component other than the standard external forward shock emission is required to account for the X-ray emission \citep{bbb+14, ewo+14, ptg+14}. \citet{ewo+14} and \citet{zs14} both argue that the late-time X-ray emission is the result of dust-scattering of the prompt emission down to X-rays, whereas \citet{bbb+14} and \citet{ptg+14} find that, similar to GRB~101225A \citep[][although see Levan et al., 2013]{tuf+11}, a blackbody component in addition to the typical power-law external shock emission component, provides a good fit to the data \citep[see also][for a two blackbody plus power-law model]{br15}. Despite differences in the interpretation of the X-ray afterglow emission mechanism, a general consensus is that the slight offset between the GRB position and the host galaxy nucleus, and more significantly, the duration of the X-ray emission, favour a collapsar rather than TDE. Furthermore, both \citet{ewo+14} and \citet{ptg+14} find that the low luminosity of the external shock emission implies a low-density circumburst environment, which would be consistent with low mass-loss expected from a BSG. However, such a progenitor model would also predict a low-metallicity environment \citep{gsa+13,nks+13}, and in the case of GRB~130925A, \citet{ptg+14} proposed a Pop III progenitor star. This is in contrast to the host galaxy observations presented here.

A BSG progenitor was also proposed in the case of GRB~111209A \citep[][although see \citealt{gmk+15}]{gsa+13,lts+14}, which similar to GRB~101225A, had a very blue and compact host galaxy. There is currently limited information published on the host galaxy properties of GRB~121027A. From HST observations, \citet{lts+14} estimate the host galaxies of GRB~101225A and GRB~111209A to have radii of $<600$~pc and $\sim 700$~pc respectively, and spectroscopic observations of the host galaxy of GRB~111209A show that it had a sub-solar metallicity in the range 12+log(O/H)=7.9-8.3 \citep{lts+14,kmf+15}. The host galaxy of GRB~130925A, on the other hand, is much larger, with an effective radius of $\sim 2.4$~kpc, and a super solar metallicity of 12+$\log$(O/H)$\sim8.8$. 

In both the case of a WR star and BSG progenitor, the hydrogen envelope has to be shed prior to the core-collapse without losing too much angular momentum. Within a more simple treatment of stellar evolution, where, for example, the rotation is modelled as a solid body, this places an upper limit on the progenitor metallicity of $Z\mathrel{\hbox{\rlap{\hbox{\lower4pt\hbox{$\sim$}}}\hbox{$<$}}}0.3Z_\odot$ \citep{hmm05,wh06}, which limits the mass-loss induced through radiation pressure. Once the progenitor star increases to above $\sim 0.5Z_\odot$, there is an excessive removal of angular momentum within the line-pressure driven stellar wind, and the blackhole-accretion disk system necessary to produce a GRB cannot form. However, in more complex models, additional considerations such as the effect of differential rotation or a companion star will also have an influence on the loss of angular momentum, and may thus relax the metallicity constraints. For example, when differential rotation is considered, it becomes possible for solar metallicity stars with initial stellar masses between 40~M$_\odot$ and 60~M$_\odot$ to collapse to form a GRB \citep{gem+12}.

A star may also retain a high angular momentum when it is part of a binary system either through tidal locking in which accretion onto the secondary star can significantly spin-up the stellar core, or through stellar core coalescence \citep{pmn+04}. Nevertheless, binary evolution models suggest that both cases are more likely to arise in low-metallicity environments, and it is still not clear whether a progenitor core could retain enough angular momentum to form a GRB. Although the core would be spun-up significantly during the helium burning phase, it is likely that a large fraction of this angular momentum would be lost prior to core-collapse. The gain in angular momentum would result in an increased stellar wind, which in turn would reduce the radius of the star and increase the binary orbit, thus resulting in a weaker tidal interaction \citep{dlp+08}. One possible solution is that the mass loss occurs preferentially along the rotational access, thus removing less angular momentum \citep{mm07}. A weak coupling between the core and envelope of the progenitor star would also alleviate this constraint, such as in the absence of magnetic fields \citep{irt04,ply+05,dlp+08}. However, these additional conditions needed to form a GRB in a solar metallicity environment add similar complexities to the binary system as in the single star collapsar model.

Owing to the very recent discovery of ultra-long GRBs, there has been little opportunity to investigate and develop possible progenitor models. Furthermore, given that GRB~130925A is the first example of an ultra-long GRB within a super-solar metallicity host galaxy, little investigation has been done on how such ultra-long duration GRBs may form within solar metallicity environments. The additional mass required to power an ultra-long GRB would imply that the conditions needed to maintain sufficient angular momentum at the time of core-collapse are likely to be even more stringent than those outlined above for standard long duration GRBs.

An alternative, model-independent reconciliation to the apparent conflict between the low metal abundances predicted by collapsar models, and observations of host galaxies with solar metallicity and above, is that the high metallicities measured in the environments of some GRBs (either from afterglow absorption lines or host galaxy emission lines) are not representative of the progenitor metallicity, and the same argument could be used in the case of the ultra-long GRB~130925A. Although we measure a super-solar metallicity both along the disk of the host galaxy, which is consistent with the afterglow position of GRB~130925A, and in the outer regions, at SF2, the host galaxy of GRB~130925A is observed edge on, and we therefore cannot rule out a strong variation in metallicity along the radial direction. If the GRB occurred at the far side of the host galaxy, it would be feasible for the GRB to have arisen in a metal-poor pocket of the host galaxy. Furthermore, HST observations of the host galaxy show evidence for an irregular galaxy morphology with a possible polar-ring structure \citep{tlh+13}. Significant variation in the metallicity distribution may thus be expected. Nevertheless, the detection of the very dusty and metal-rich host galaxy of GRB~130925A so early on after the discovery of the first ultra-long GRB would suggest that such environmental conditions are not rare.

Although there is substantial evidence indicating that long GRBs (and possibly also the ultra-long class) are more likely to arise in sub-solar metallicity environments with $z<0.5Z_\odot$ \citep[especially noticeable at $z\mathrel{\hbox{\rlap{\hbox{\lower4pt\hbox{$\sim$}}}\hbox{$<$}}} 1.5$; e.g.][]{mkk+08,lkb+10,gf13,kmf+15}, it is clear that long GRBs do not adhere to a strict metallicity cut-off. It is, therefore, important to consider the effect that other properties apart from metallicity have on the loss of angular momentum over a stars lifetime \citep{lsk+10}, such as anisotropic mass-loss \citep{mm07}, wind clumping \citep{cdh+02}, and magnetic processes \citep{dbl+08}. In order to compare observations with progenitor model predictions, it would be beneficial to quantify the probability that any particular progenitor model has in evolving to form a GRB \citep[e.g.][]{hmm05,hp13}, together with the observable signatures of any given model, wherever possible. This would then permit a direct comparison between the probability distributions and e.g. the observed metallicity distribution. However, ultimately, larger samples of spatially resolved GRB host galaxies are needed, and especially for host galaxies with super-solar metallicity, so that variations in the environmental conditions within the vicinity of the GRB relative to the rest of the host galaxy can be assessed, and underlying common properties can be identified.

\section{Summary}
In this paper we have presented a detailed analysis on the spectroscopic X-shooter observations taken of the host galaxy of the ultra-long duration GRB~130925A (lasting for $\sim 20$~ks). The position of the slit fell along the disk of the galaxy, and fortuitously crossed the GRB explosion sight, close to the nucleus of the host galaxy, and a star forming region located in the outskirts of the galaxy. Our observations suggest that the GRB occurred in a metal-rich region of the host galaxy, with a higher dust content than that of a neighbouring HII region, which is in apparent conflict with the favoured progenitor models of both standard long duration GRBs, and the more recently discovered but more poorly understood class of ultra-long GRBs. 

Although the sample of long GRBs with dust-rich and super-solar metallicity host galaxies has drastically increased over the past half a decade, our spectroscopic data of GRB~130925A are only the second such example of spatially resolved observations of a super-solar GRB host galaxy, and they are the first such observations for an ultra-long GRB. The edge-on orientation of the host galaxy of GRB~130925A, and the very central sightline to the GRB through the galaxy disk, raises the possibility that GRB~130925A occurred within a metal-poor region, on the far side of the host galaxy. Nevertheless, the increased range in observed host galaxy properties provided by observations such as those presented here, and especially at such a relatively low redshift, provides greater constraint on the GRB progenitor models, and ultimately, on the environmental properties traced by GRBs. 

With the currently small sample of ultra-long GRBs, the statistics for this class of event remain too small to draw any strong conclusions on the likely progenitors producing these incredibly long-lived and energetic events. Nevertheless, observations such as our X-shooter data of the host galaxy of GRB~130925A, present new challenges to the favoured models. To continue in this vein it is thus important to be mindful of the dominant selection effects present in GRB and host galaxy samples, of which dust-extinction is possibly the most dominant, and to pursue observing programmes that address these selection effects, thus enabling us to quantify the extent of our biases.

\begin{acknowledgements}
P.S. acknowledges support through the Sofja Kovalevskaja Award from the Alexander von Humboldt Foundation of Germany. J.F.G., M.T. and P.W. acknowledge support through the Sofja Kovalevskaja Award to P.~Schady from the Alexander von Humboldt Foundation of Germany. S.K. and A.N.G. acknowledge support of DFG grant KI 766/16-1. C.D. acknowledges support through EXTraS, funded from the European Union's Seventh Framework Programme for research, technological development and demonstration under grant agreement no. 60752. S.S. acknowledges support by the Th{\"u}ringer Ministerium f{\"u}r Bildung, Wissenschaft ind Kultur under FKZ 12010-514. D.A.K. acknowledges TLS Tautenburg for financial support. Part of the funding for GROND (both hardware as well as personel) was generously granted from the Leibniz-Prize to Prof.~G.~Hasinger (DFG grant HA 1850/28-1). This study is based on data acquired at ESO, Programme ID 091.A-0703.
\end{acknowledgements}

\end{document}